\documentclass[structabstract]{aa}

\usepackage{txfonts}
\usepackage{graphicx}
\usepackage{natbib}
\usepackage{color}
\bibpunct{(}{)}{;}{a}{}{,}

\newcommand{\stat}{    {Ann. Stat.}}
\newcommand{\asr}{    {Adv. Space Res.}}

\titlerunning{Wave properties of CBFs}
\authorrunning{Long et al.}

\begin{document}

\title{Deceleration and Dispersion of Large-scale Coronal Bright Fronts}

\author{David~M.~Long\inst{\ref{inst1},\ref{inst2}} \and Peter~T.~Gallagher\inst{\ref{inst1}} \and R.~T.~James~McAteer\inst{\ref{inst1},\ref{inst3}} \and D.~Shaun~Bloomfield\inst{\ref{inst1}}}

\institute{Astrophysics Research Group, School of Physics, Trinity College Dublin, Dublin 2, Ireland. \email{longda@tcd.ie}\label{inst1}
\and Harvard-Smithsonian Centre for Astrophysics, 60 Garden St., Cambridge, MA 02138, USA\label{inst2}
\and Department of Astronomy, New Mexico State University, P.O. Box 30001 MSC 4500, Las Cruces, NM 88003, USA\label{inst3}}

\abstract{One of the most dramatic manifestations of solar activity are large-scale coronal bright fronts (CBFs) observed in extreme ultraviolet (EUV) images of the solar atmosphere. To date, the energetics and kinematics of CBFs remain poorly understood, due to the low image cadence and sensitivity of previous EUV imagers and the limited methods used to extract the features.}{In this paper, the trajectory and morphology of CBFs was determined in order to investigate the varying properties of a sample of CBFs, including their kinematics and pulse shape, dispersion, and dissipation.}{We have developed a semi-automatic intensity profiling technique to extract the morphology and accurate positions of CBFs in 2.5--10 min cadence images from \emph{STEREO}/EUVI. The technique was applied to sequences of 171~\AA\ and 195~\AA\ images from \emph{STEREO}/EUVI in order to measure the wave properties of four separate CBF events.}{Following launch at velocities of $\sim$240--450~km~s$^{-1}$ each of the four events studied showed significant negative acceleration ranging from $\sim -$290 to $-$60~m~s$^{-2}$. The CBF spatial and temporal widths were found to increase from $\sim$50~Mm to $\sim$200~Mm and $\sim$100~s to $\sim$1500~s respectively, suggesting that they are dispersive in nature. The variation in position-angle averaged pulse-integrated intensity with propagation shows no clear trend across the four events studied. These results are most consistent with CBFs being dispersive magnetoacoustic waves.}{}


\maketitle

\section{Introduction}
\label{sect:intro}

``EIT waves'' were first observed in the solar corona using the \emph{SOlar and Heliospheric Observatory} (\emph{SOHO})/Extreme ultraviolet Imaging Telescope \citep[EIT;][]{Moses:1997vn} and analysed in detail by \citet{Thompson:1998ab}. They have been a source of much controversy and debate in the solar physics community since this initial observation, with different authors suggesting that they are alternatively fast-mode MHD waves \citep[e.g.:][]{Wang:2000tg,Warmuth:2004rm,Long:2008eu,Veronig:2008ud,Gopalswamy:2009dn}, a result of the re-structuring of the magnetic field during the eruption of a coronal mass ejection \citep[CME;][]{Chen:2002rw,Chen:2005xe,Attrill:2006vn,Attrill:2007vn,Delannee:2007kx,Delannee:2008uq} or a coronal MHD soliton \citep{Wills-Davey:2007oa}. It should be noted at this point that the name ``EIT wave'' is typically used for historical reasons. To reflect the uncertainty surrounding the physical interpretation of ``EIT waves'' we shall adopt the suggested nomenclature of \citet{Gallagher:2010ab} and refer to these disturbances as coronal bright fronts (CBFs) throughout this paper.

The uncertainty in the nature of CBFs has arisen from conflicting results being drawn from the same observations. A pseudo-wave interpretation was proposed following observations of stationary bright fronts \citep{Delannee:1999ab}, a strong correlation with CMEs \citep{Biesecker:2002lq} and a lower than expected estimated pulse velocity \citep{Wills-Davey:2007oa}. However, observations of refraction \citep{Wang:2000tg,Ofman:2002ab,Veronig:2006fy} and reflection \citep{Gopalswamy:2009dn} of CBFs at coronal hole boundaries would appear to suggest a wave interpretation. A conclusive result has been hampered by the diffuse nature of CBFs and the relatively low temporal cadence of the observing instruments, both of which make it difficult to characterize their true nature. A full review of CBFs, their morphology, kinematics, relationship to other solar phenomena, and theoretical interpretations may be found in \citet{Wills-Davey:2010ab} and \citet{Gallagher:2010ab}. 

The first observations of CBFs were made using EIT with an effective cadence of $\sim$12~minutes in the 195~\AA\ passband. Initial estimates of the kinematics of these disturbances, using a point-and-click methodology applied to running-difference images, estimated the average velocity at $\sim$189~km~s$^{-1}$ \citep{Thompson:2009yq}. A higher velocity of $311\pm111$~km~s$^{-1}$ was found by \citet{Warmuth:2004rm} using additional passbands to compensate for the lack of 195~\AA\ images. This is comparable to the range of Alfv\'{e}n speeds predicted by \citet[][$\sim$215--1500~km~s$^{-1}$]{Wills-Davey:2007oa}. The \emph{Solar TErrestrial RElations Observatory} \citep[\emph{STEREO};][]{Kaiser:2008ab} mission with the Extreme UltraViolet Imager \citep[EUVI;][]{Wuelser:2004bs} instrument has led to new results. EUVI has an effective observing cadence of up to 1.5~minutes (ten times that of EIT), allowing for an improved estimate of the kinematics of these disturbances. A numerical differencing technique was applied by \citet{Long:2008eu} to running-difference EUVI images to estimate a peak velocity range for a CBF of $\sim$153 to 475~km~s$^{-1}$ with the acceleration of the disturbances estimated to be between $-413$ and 816~m~s$^{-2}$, depending on the cadence of the observations. For the same event, \citet{Veronig:2008ud} estimated the CBF velocity to be 460~km~s$^{-1}$ with an associated deceleration of $-160$~m~s$^{-2}$ by fitting a quadratic model to distance-time measurements.  The results of \citet{Long:2008eu} also indicated that the lower cadence of \emph{SOHO}/EIT had resulted in the kinematics of CBFs being previously underestimated, an observation confirmed by \citet{Veronig:2008ud} and \citet{Ma:2009fk}. 

The kinematics of CBFs provide an insight into the true nature of the disturbances. Another useful physical indicator is the presence of pulse broadening, previously described by \citet{Warmuth:2004ab}. Several authors have examined CBF pulse broadening using observations of multiple events from different passbands, including a combination of EUV and H$\alpha$ observations \citep{Warmuth:2001ab, Warmuth:2010ab,Veronig:2010ab} as well as observations of a disturbance in soft X-ray data from \emph{GOES}/SXI \citep{Warmuth:2005vf}. In contrast, \citet{Wills-Davey:2006ab} observed no measurable increase in the FWHM of a pulse using high-cadence observations of a single CBF across the limited field-of-view of the \emph{Transition Region And Coronal Explorer} (\emph{TRACE}). This led \citet{Wills-Davey:2007oa} to propose that CBFs were soliton-like waves which exhibit no significant dispersion with propagation. 

Observations of CBFs typically show a decreasing front intensity with propagation. This has been noted by many authors including \citet{Warmuth:2001ab, Warmuth:2004ab} and \citet{Veronig:2010ab}, and is interpreted as evidence of energy conservation as a wave expands into a larger area. The specific geometry of the wave expansion can be derived by measuring the decay in pulse intensity and growth of pulse width, although previous efforts to measure these parameters accurately have been hampered by the small number of data points associated with each observed event \citep{Warmuth:2004ab}. 

In this paper we determine the kinematics of several CBFs and examine the variation in their width and intensity with increasing time (and hence distance). The data are presented and discussed in Section~\ref{sect:observations}, with Section~\ref{sect:methods} detailing the analysis method. The results are discussed in Section~\ref{sect:results} and some conclusions drawn in Section~\ref{sect:conclusion}, along with some thoughts on future work.

\section{Observations}
\label{sect:observations}

The data discussed here were obtained using EUVI, part of the Sun Earth Connection Coronal and Heliospheric Investigation \citep[SECCHI;][]{Howard:2008fq} suite of instruments onboard the \emph{STEREO}-A and \emph{STEREO}-B spacecraft. EUVI is a normal-incidence telescope of Ritchey-Chr\'{e}tien design with a pixel scale of 1.6\arcsec. It observes the Sun in four passbands (304~\AA, 171~\AA, 195~\AA, and 284~\AA), with the peak temperature sensitivities for each passband being approximately 0.07~MK (304~\AA), 1~MK (171~\AA), 1.5~MK (195~\AA), and 2.25~MK (284~\AA). The imaging cadence of the EUVI instrument ranges from 1.5~minutes (in the 171~\AA\ passband) to 20~minutes (in the 284~\AA\ passband).

\begin{figure}[t]
\centering{
               \includegraphics[width=0.5\textwidth,clip=,trim=0mm 0mm 0mm 5mm]{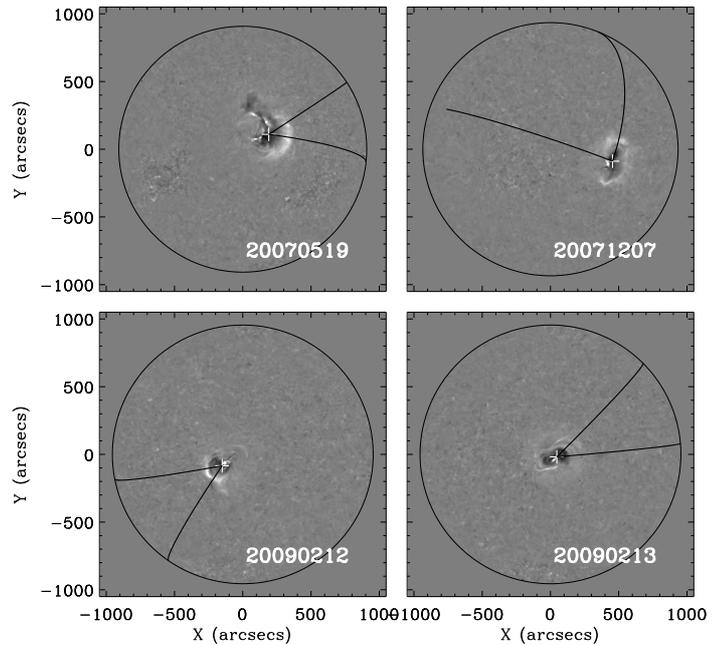}
               }
\caption{Percentage base difference images of the 2007~May~19 (top left), 2007~Dec~07 (top right), 2009~Feb~12 (bottom left) and 2009~Feb~13 (bottom right) CBFs, each in the 195~\AA\ passband as seen by \emph{STEREO}-B. The solid lines indicate the region in which the disturbance was identified using the intensity profile technique, while the cross marks the estimated origin of the event. White (black) is an increase (decrease) in intensity from the base image.}
\label{fig:image}
\end{figure}

Although EUVI takes observations in all four of these passbands, only the 171~\AA\ and 195~\AA\ passbands were used here. This is due to the high temporal cadence of both passbands (1.5--2.5 minutes for 171~\AA\ and 5--10 minutes for 195~\AA) and also as CBFs are more readily observed (i.e.,\ of higher contrast) in these two passbands. CBFs have been observed in the 284~\AA\ passband \citep{Zhukov:2004kh} and the 304~\AA\ passband \citep{Long:2008eu}, but the nature of the data make it difficult to use these passbands for more rigorous analysis. 

The CBFs were studied using de-rotated base-difference (BD) and percentage base-difference (PBD) images. This involved de-rotating all images for a given event to the same pre-event time, in order to correct for solar rotation between images, and then subtracting a pre-event image to produce a BD image. The ratio of the BD image and the pre-event image was then calculated, giving a PBD image. This is described by the equation,
\begin{equation}
I_{PBD} = \frac{I_t - I_0}{I_0} \times 100  \  ,
\end{equation}
where $I_t$ is the image at any time $t$ and $I_0$ is the pre-event image. This technique produces images that highlight the CBF and any associated dimming regions, with the intensity values of any given pixel corresponding to the percentage change in intensity with respect to the pre-event image \citep[for more details see][]{Wills-Davey:2007oa}. 

Figure~\ref{fig:image} shows PBD images for the four events studied here from \emph{STEREO}-B, with the event dates indicated in the lower right of each panel. The erupting AR was close to disk centre for both spacecraft for both the 2007~May~19 and 2007~Dec~07 events due to their small separation. However the eruption was close to disk centre for \emph{STEREO}-B but on the limb for \emph{STEREO}-A for both the 2009~Feb~12 and 2009~Feb~13 events. As a result, only the \emph{STEREO}-B observations are considered here. The observed pulse in each case extended over a large fraction of the solar disk and was observed in multiple images from both passbands. The observing cadence of the 195~\AA\ passband was 600~s for each event, while the 171~\AA\ passband operated at a cadence of 150~s for each event with the exception of the 2009~Feb~13 event where it had a cadence of 300~s.

\section{Methods}
\label{sect:methods}

\begin{figure*}
\sidecaption
               \includegraphics[width=0.67\textwidth,clip=,trim=0mm 5mm 0mm 0mm]{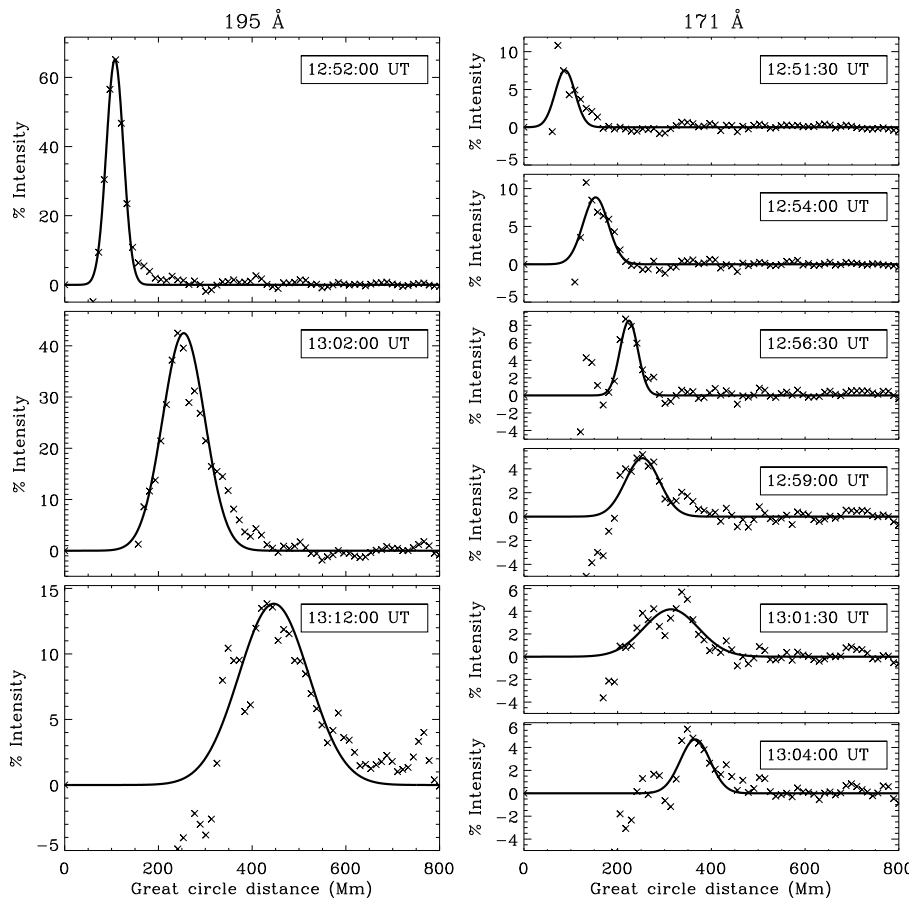}
\caption{PBD image intensity profiles (crosses) for the 2007~May~19 event as obtained from \emph{STEREO}-A in the 195~\AA\ (left) and 171~\AA\ (right) passbands. The Gaussian fit (solid curve) to the positive section of each profile has been overplotted on each panel. The time of the leading image ($I_t$) in each case is on the upper right of the panel. More cases are displayed in the online Figures~\ref{fig:profile_20070519_B} to \ref{fig:profile_20090213_B}.}
\label{fig:profile}
\end{figure*}

\onlfig{3}{
\begin{figure*}
\sidecaption
              \includegraphics[width=0.67\textwidth,clip=,trim=0mm 5mm 0mm 0mm]{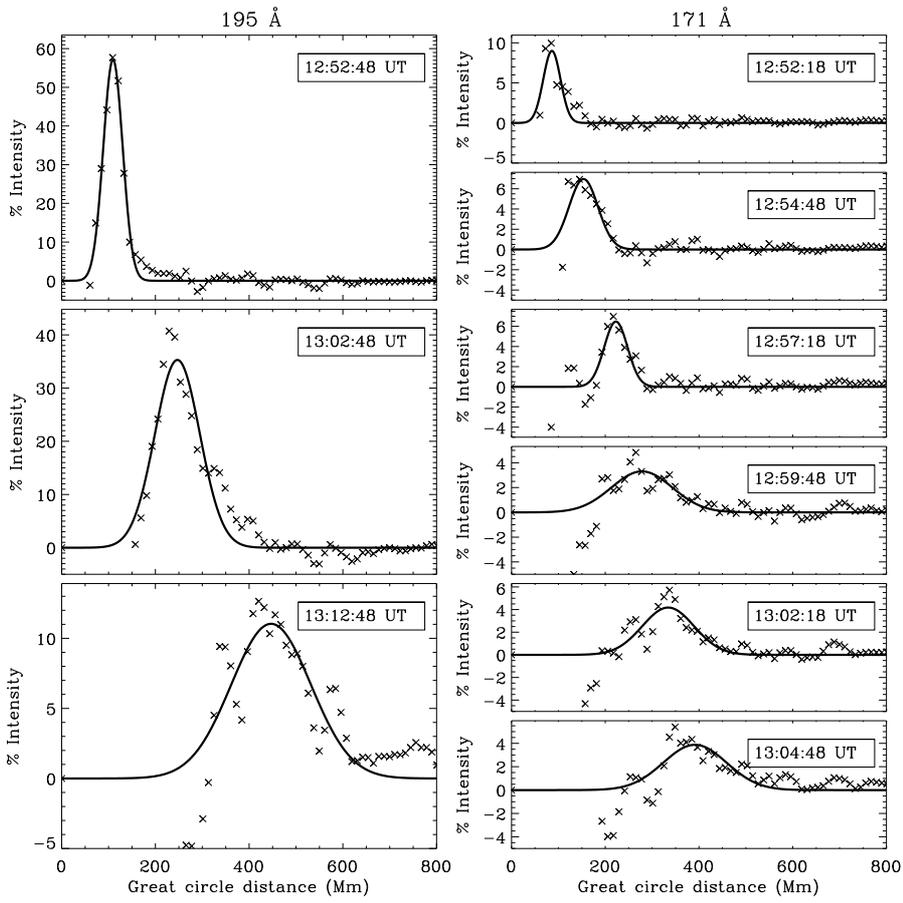}
\caption{Same as for Figure~\ref{fig:profile} but for 2007~May~19 \emph{STEREO}-B.}
\label{fig:profile_20070519_B}
\end{figure*}
}
\onlfig{4}{
\begin{figure*}
\sidecaption
              \includegraphics[width=0.67\textwidth,clip=,trim=0mm 5mm 0mm 0mm]{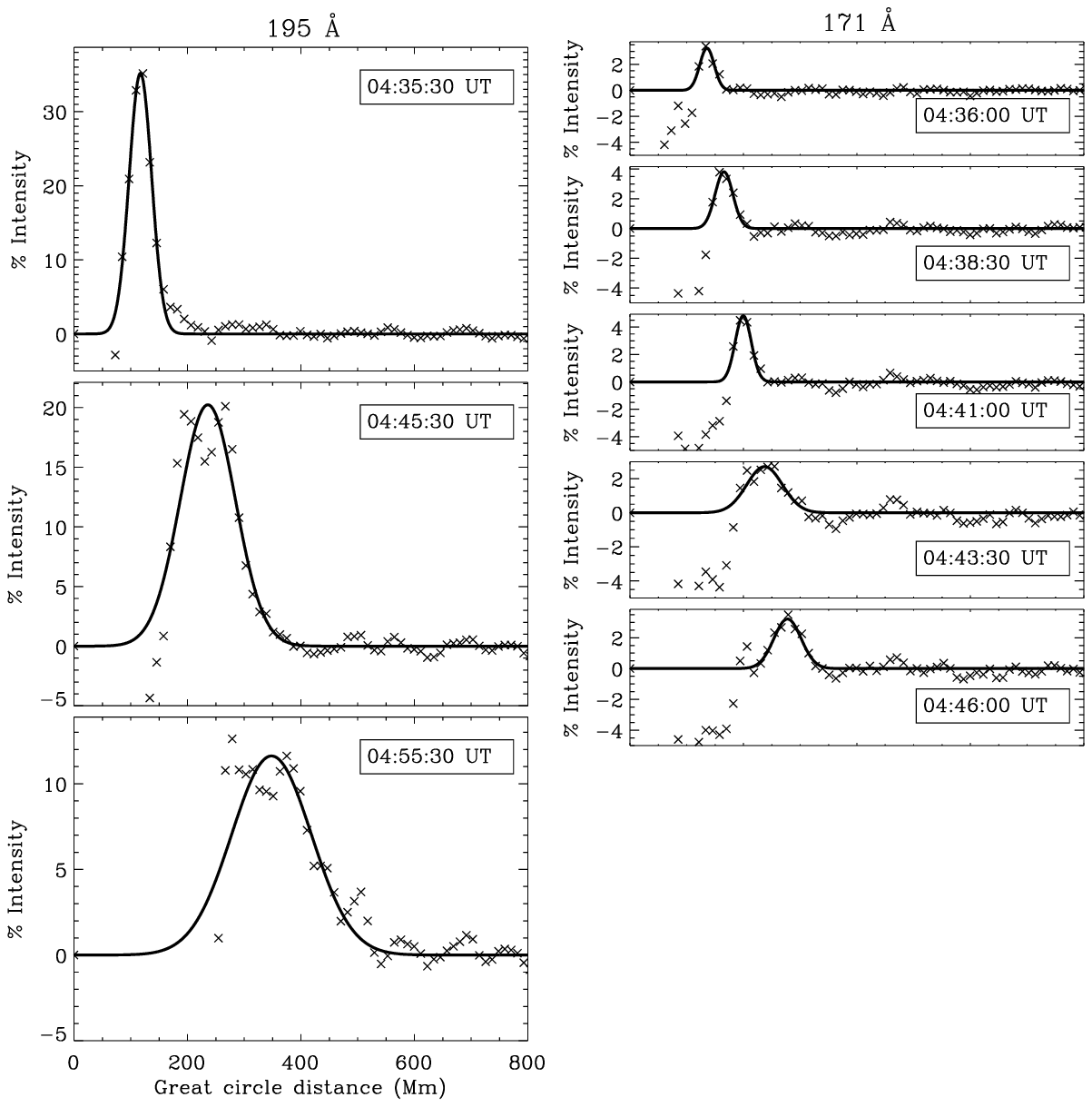}
\caption{Same as for Figure~\ref{fig:profile} but for 2007~Dec~07 \emph{STEREO}-A.}
\label{fig:profile_20071207_A}
\end{figure*}
}
\onlfig{5}{
\begin{figure*}
\sidecaption
              \includegraphics[width=0.67\textwidth,clip=,trim=0mm 5mm 0mm 0mm]{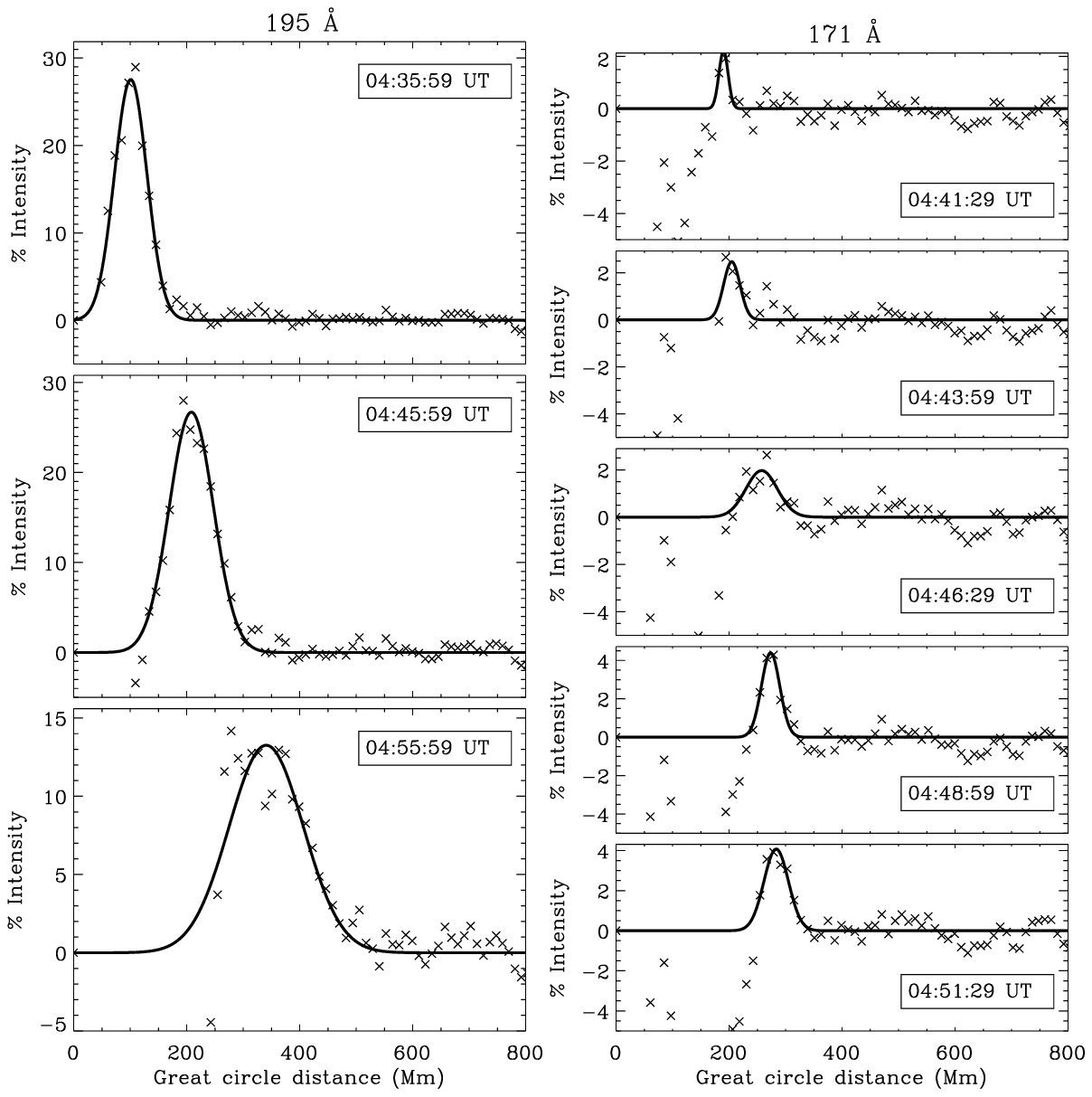}
\caption{Same as for Figure~\ref{fig:profile} but for 2007~Dec~07 \emph{STEREO}-B.}
\label{fig:profile_20071207_B}
\end{figure*}
}
\onlfig{6}{
\begin{figure*}
\sidecaption
              \includegraphics[width=0.67\textwidth,clip=,trim=0mm 5mm 0mm 0mm]{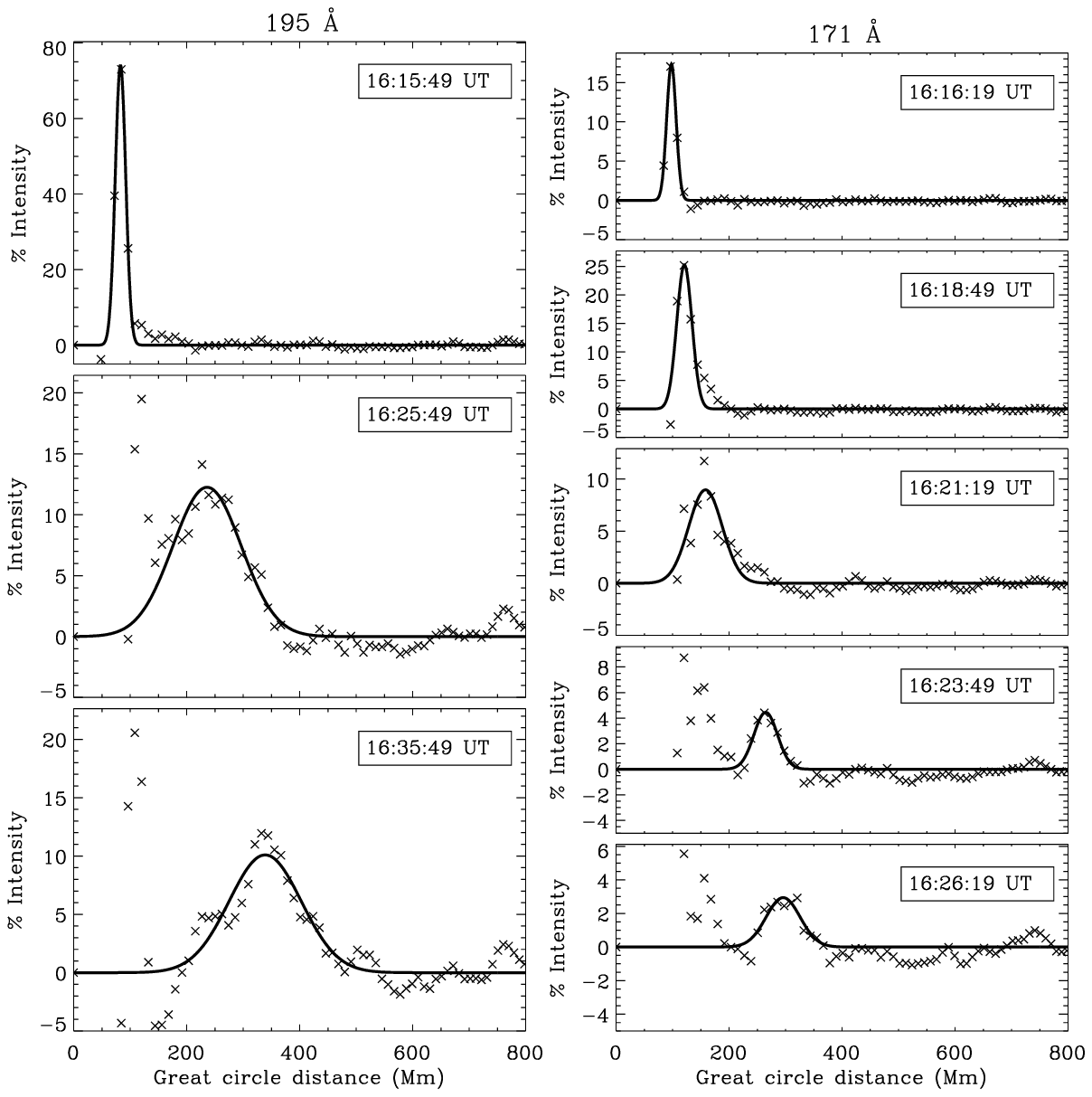}
\caption{Same as for Figure~\ref{fig:profile} but for 2009~Feb~12 \emph{STEREO}-B.}
\label{fig:profile_20090212_B}
\end{figure*}
}
\onlfig{7}{
\begin{figure*}
\sidecaption
              \includegraphics[width=0.67\textwidth,clip=,trim=0mm 5mm 0mm 0mm]{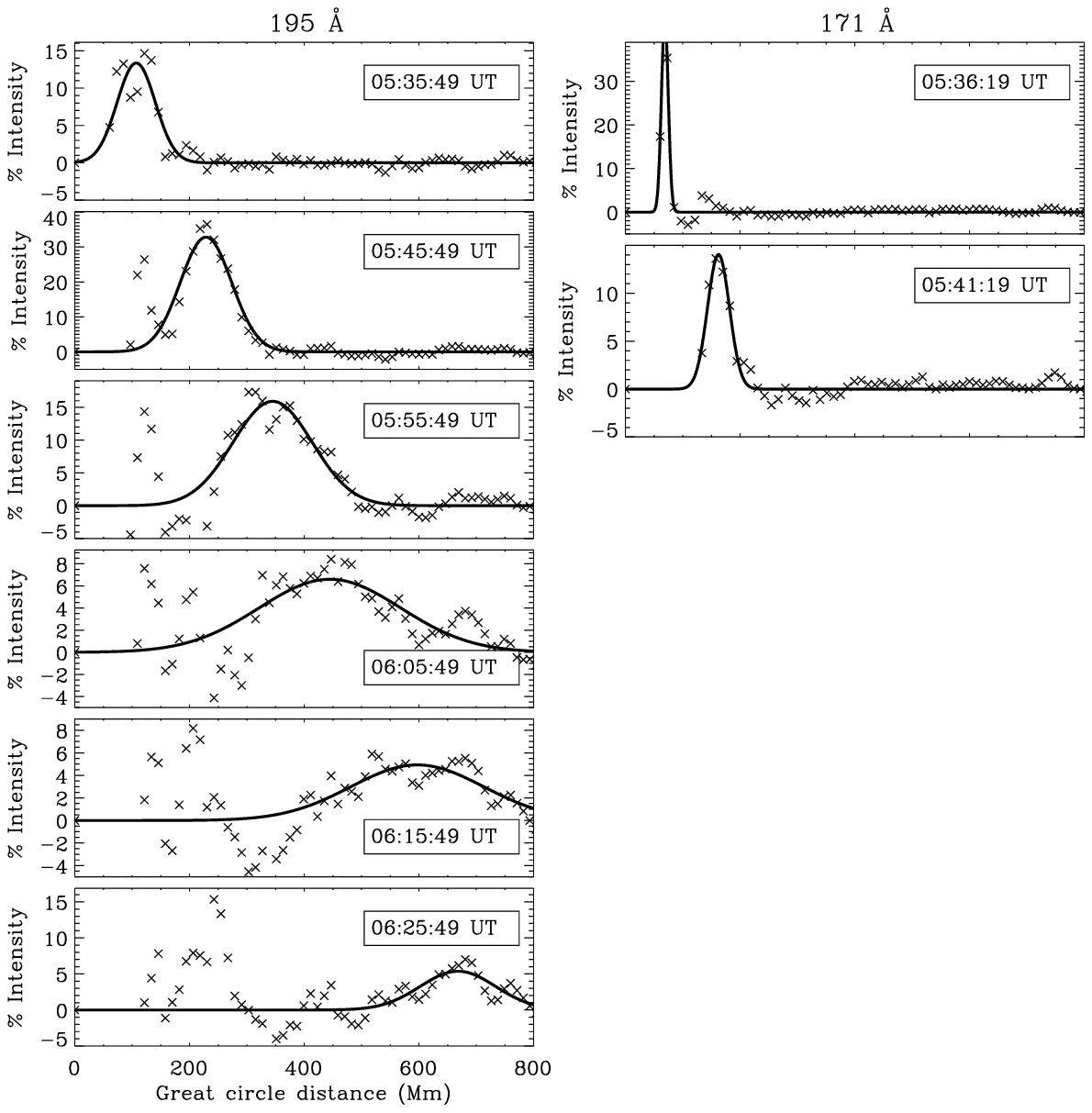}
\caption{Same as for Figure~\ref{fig:profile} but for 2009~Feb~13 \emph{STEREO}-B.}
\label{fig:profile_20090213_B}
\end{figure*}
}

It is well known that the use of point-and-click techniques in conjunction with PBD images results in large errors that are unique to each user. As a result, we developed an algorithm to automatically identify the presence of CBFs in PBD images and produce estimates of the kinematics of these CBFs. This algorithm is outlined in Section~\ref{subsect:identification}, with the resulting kinematics presented in Section~\ref{subsect:kinematics}. The algorithm was also used to investigate the variation in both the width (Section~\ref{subsect:broadening}) and position angle (PA) -averaged integrated intensity (Section~\ref{subsect:intensity}) of the CBF pulse.

\subsection{Pulse Identification}
\label{subsect:identification}

The source of the CBF disturbance was identified from the PBD images by fitting ellipses to the visually traced wavefronts from the first two observations of the disturbance in both the 171~\AA\ and 195~\AA\ passbands, with the mean of the centres of the four ellipses taken as the origin of the CBF and the standard deviation giving the associated error (typically $\sim$20~Mm). Next, a great-circle sector (i.e., an area on the sphere bounded by two great circles) projected onto the Sun was identified in which the disturbance was clearly visible throughout its propagation, with the source of the disturbance acting as the crossing point of the two bounding great circles (see Figure~\ref{fig:image}). From each image the PBD intensity within this sector was then averaged across PA, in annuli of increasing radii with 1~degree width on the surface of the sphere, to produce an intensity profile as a function of distance away from the source location \citep[cf. similar techniques proposed by][]{Warmuth:2004ab,podladchikova2005,Wills-Davey:2006ab,Veronig:2010ab}. This process was repeated to create BD intensity profiles as well.

The intensity profile produced by this technique was fitted using a Gaussian model because CBFs were noted by \citet{Wills-Davey:2006ab} to have a Gaussian form. The technique may be viewed in the attached movie (see the movie attached to the online Figure~\ref{fig:movie_cover}), while the resulting intensity profiles for the 2007~May~19 event as obtained by \emph{STEREO}-A in the 171~\AA\ and 195~\AA\ passbands are displayed in Figure~\ref{fig:profile}. Note that the first three observations in the 195~\AA\ passband and the first six observations in the 171~\AA\ passband from both spacecraft were used for this work. Beyond this, the $\chi^2$ values of the Gaussian fits were too large for accurate analysis. 

\onlfig{8}{
\begin{figure*}
\centering{
              \includegraphics[width=0.9\textwidth,clip=,trim=0mm 5mm 0mm 70mm]{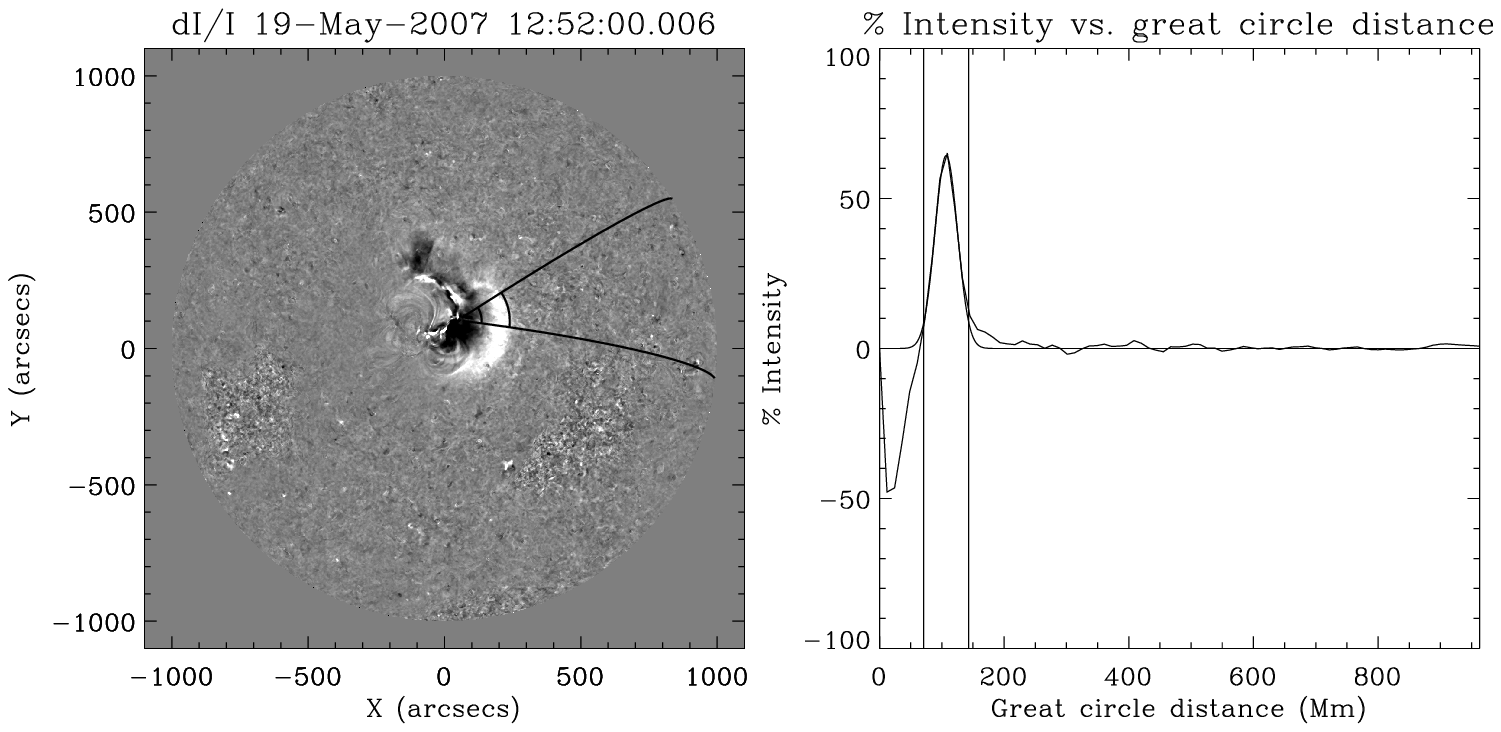}
              }
\caption{\emph{Left panel}: PBD image with arc sector over-plotted. \emph{Right panel}: Intensity profile across arc sector. The intensity profile at a given distance corresponds to the mean intensity across the arc sector at that distance. This is then fitted using a Gaussian model and repeated for the next image. The vertical lines in the right-hand plot and arc lines in the left-hand plot correspond to the $2\sigma$ limits of the Gaussian fit. The attached movie shows the procedure used to build these diagrams.}
\label{fig:movie_cover}
\end{figure*}
}

The intensity profile technique is judged to be more accurate than a normal point-and-click technique as it is semi-automated and reproducible, with any associated errors quantifiable as they result from the fitting of a Gaussian to an intensity profile across the pulse. In contrast, the point-and-click methodology commonly used to identify CBFs is highly user-dependent, with large unquantifiable errors in the identification of the pulse position. Although a pulse may be tracked over larger distances using a visual method, increased image processing is often required to improve the visibility of the pulse. However, the intensity profile technique used here applies the same processing for all images, with the result that measurements of the pulse are directly comparable. The intensity profile technique can also be used to process large amounts of data (such as those from \emph{SDO}) rapidly as it does not require the same degree of user input as the visual method.

In comparing different analysis techniques, it was also noted that BD and PBD images produced a more accurate estimate of the location of the pulse than running-difference images as they highlight relative motion of solar features with respect to a pre-event image rather than the previous image. This means that the bright pulse indicates the entirety of the CBF, rather than the portion of the pulse which has moved beyond its previous extent \citep[for a discussion of the problems associated with running-difference images see][]{Attrill:2010ab}. 

\subsection{Pulse Kinematics}
\label{subsect:kinematics}

\begin{figure}
\centerline{
    \includegraphics[width=0.45\textwidth,clip=,trim=4mm 7mm 60mm 72mm]{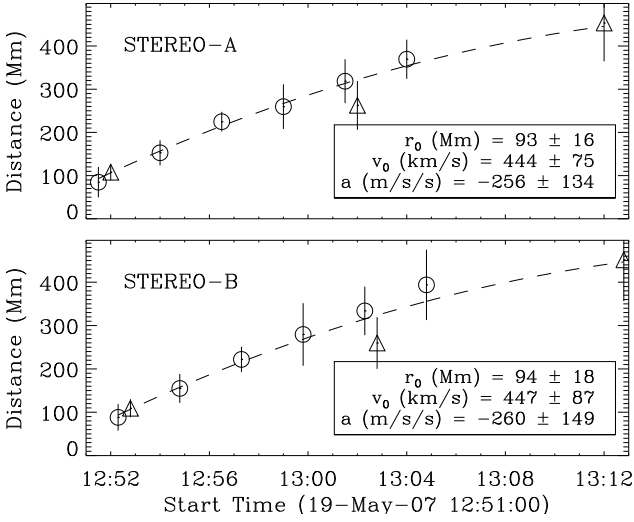}
}
\caption{Distance-time plots for \emph{STEREO}-A (top) and \emph{STEREO}-B (bottom). The 171~\AA\ (circles) and 195~\AA\ (triangles) data have been combined as they follow similar kinematical curves. The mean offset distance, initial velocity, and acceleration terms resulting from the bootstrapping analysis (fit indicated by dashed line) are also stated in the bottom right of each panel, with errors represented by the standard deviation. The errors on each point are given by the error on the mean of the Gaussian fit applied to the intensity profile. More cases are displayed in the online Figures~\ref{fig:app_kinematics_20071207} and \ref{fig:app_kinematics_200902}.}
\label{fig:kinematics}
\end{figure}

\onlfig{10}{
\begin{figure*}
\centerline{
    \includegraphics[width=0.45\textwidth,clip=,trim=4mm 7mm 57mm 72mm]{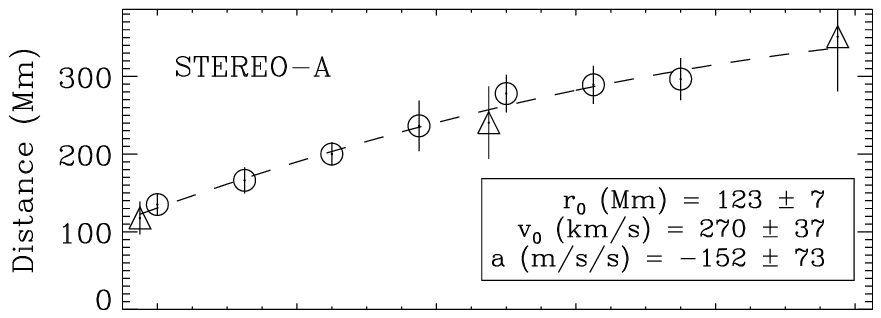}
}
\caption{Same as Figure~\ref{fig:kinematics} but for the event on 2007~Dec~07.}
\label{fig:app_kinematics_20071207}
\end{figure*}
}
\onlfig{11}{
\begin{figure*}
\centerline{
    \includegraphics[width=0.45\textwidth,clip=,trim=4mm 7mm 57mm 106mm]{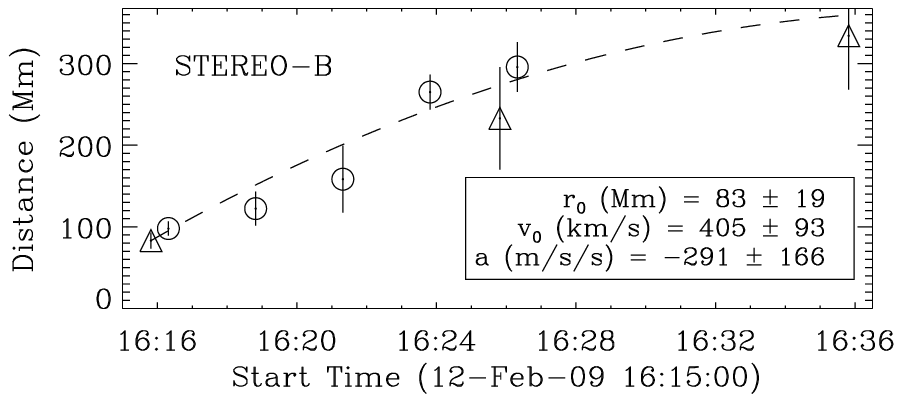}
}
\centerline{
    \includegraphics[width=0.45\textwidth,clip=,trim=4mm 7mm 57mm 106mm]{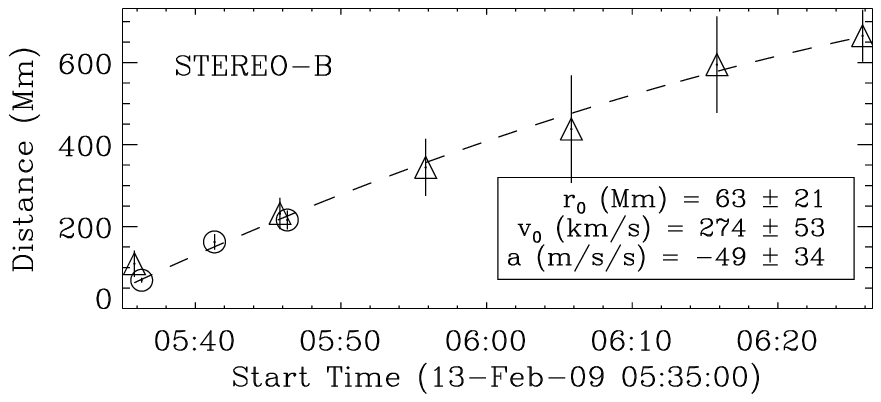}
}
\caption{Same as Figure~\ref{fig:kinematics} but for the events on 2009~Feb~12 (top) and 2009~Feb~13 (bottom), both showing \emph{STEREO}-B data.}
\label{fig:app_kinematics_200902}
\end{figure*}
}

The kinematics of the CBF pulse were studied by plotting the variation with time in the centroid of the Gaussian model applied to the positive section of the intensity profiles (see Figure~\ref{fig:profile}). The errors associated with each data-point are given by the error associated with the mean of the Gaussian fit to the intensity profile in each case. Data from both the 171~\AA\ and 195~\AA\ passbands were combined in this case as they have been observed to follow similar kinematical curves \citep{Warmuth:2004rm,Veronig:2010ab}. This was carried out for both \emph{STEREO} spacecraft (see Figure~\ref{fig:kinematics}) with the resulting plots fitted using a quadratic model of the form,
\begin{equation}
r(t) = r_0 + v_0 t + \frac{1}{2}a t^2,
\end{equation}
where $r_0$ is the offset distance, $v_0$ is the initial velocity, $a$ is the constant acceleration of the pulse, and $t$ is the time elapsed since the first observation of the disturbance. The data were analysed using a residual resampling bootstrapping technique to ensure that the results were as statistically rigorous as possible. This technique works by fitting the given data ($y_i$; $i=1, 2, \ldots, n$) with a specified model, yielding the fitted values ($\hat{y_{i}}$) and residuals ($\hat{\epsilon}=y_{i} - \hat{y_{i}}$). The residuals are then randomly ordered, randomly assigned a sign ($1$ or $-1$), and added to the original fit values to produce a new data set. These data are fit using the same model with the resulting re-fitted parameters recorded. The process of randomizing residuals, applying them to the original fit and refitting is then repeated a large number of times ($\sim$10\,000). This technique is statistically rigorous and produces a more accurate result than a simple model-fit to the given data \citep{Efron:1979ly}, allowing each fit parameter to be characterised by a distribution. 

The inset values in Figure~\ref{fig:kinematics} are the mean and standard deviation for these bootstrapped parameter distributions, which suggest that the pulse has an initial velocity of $\sim$446~km~s$^{-1}$ with a negative acceleration observed in both spacecraft ($a\simeq -260\pm130$~m~s$^{-2}$ in \emph{STEREO}-A and $a\simeq -260\pm150$~m~s$^{-2}$ in \emph{STEREO}-B). The errors associated with the acceleration terms are quite large (in each case the acceleration is negative within the 1-sigma error range). However, the kinematics are consistent between both \emph{STEREO} spacecraft, suggesting that the deceleration is real. 

The initial velocity values given here are similar to previous estimates obtained by \citet{Long:2008eu} and \citet{Veronig:2008ud} who both studied this event, although in both of those cases a three-point Lagrangian interpolation technique was used to determine the kinematics of the CBF. Although that technique retains all of the data points through the use of an interpolation method, it has been observed to introduce artificial trends through the skewing of the interpolated edge points. As a result, it is possible to misinterpret the derived velocity and acceleration plots. In contrast, the bootstrapping technique used here is designed to determine the best-fit of a model to a given small data set, producing accurate estimates of the model parameters and quantifiable associated errors.

\subsection{Pulse Width}
\label{subsect:broadening}

\begin{figure*}
\centerline{
   \includegraphics[width=0.45\textwidth,clip=,trim=0mm 6mm 55mm 93mm]{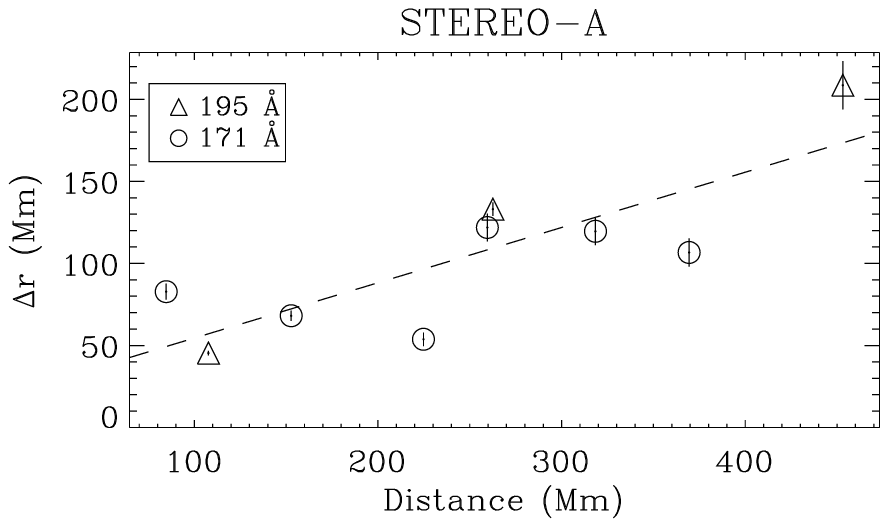}
   \includegraphics[width=0.45\textwidth,clip=,trim=0mm 6mm 55mm 93mm]{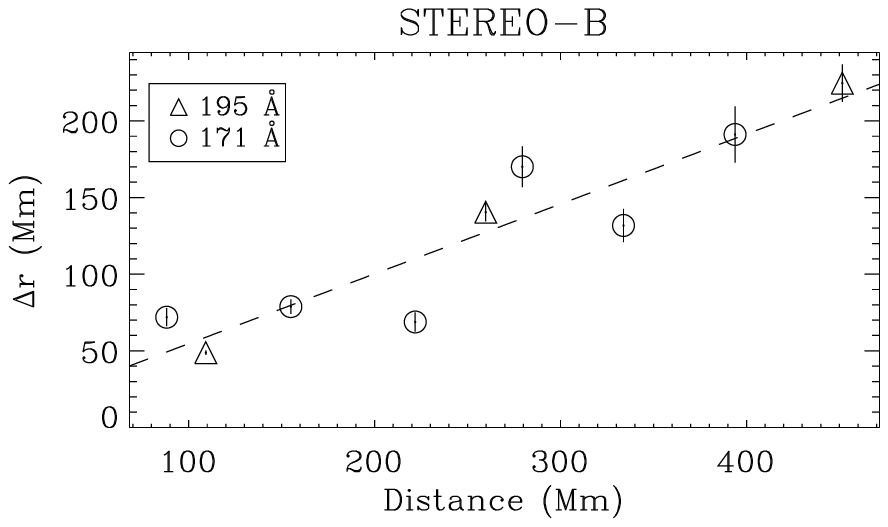}
               }
\centerline{
    \includegraphics[width=0.45\textwidth,clip=,trim=0mm 6mm 55mm 98mm]{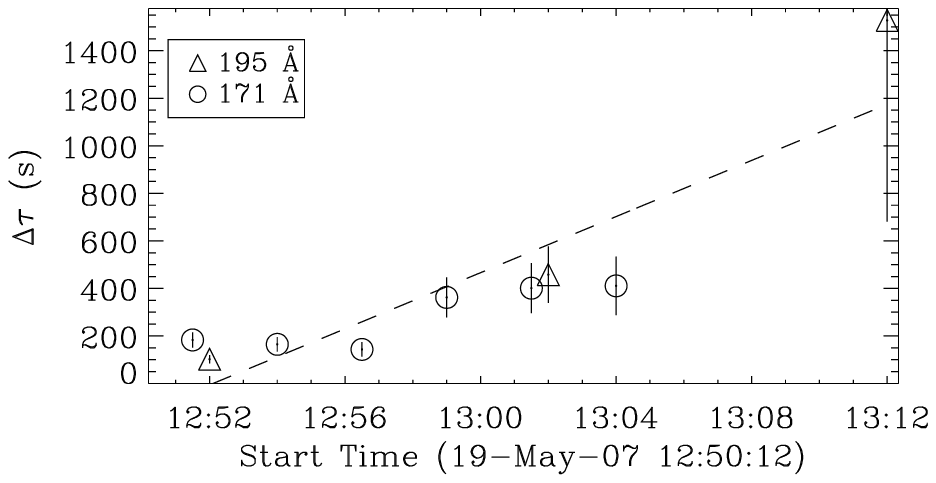}
    \includegraphics[width=0.45\textwidth,clip=,trim=0mm 6mm 55mm 98mm]{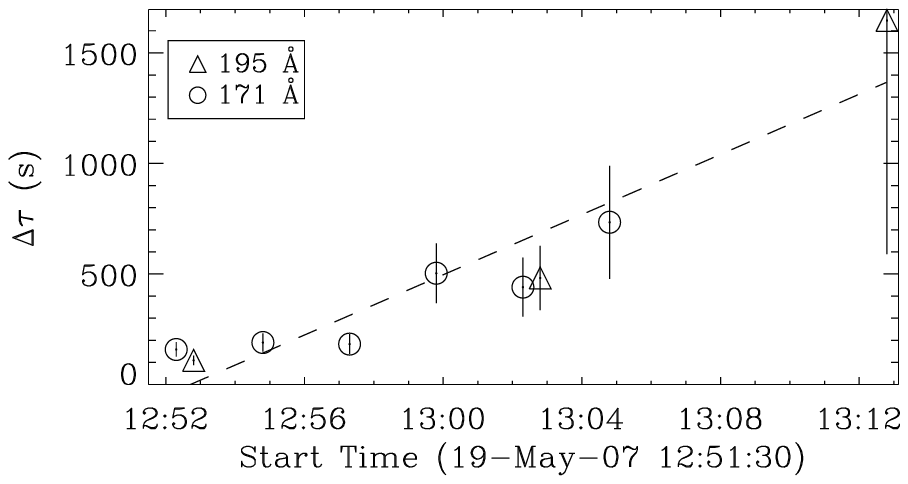}
               }
\caption{\emph{Top}: Variation in pulse spatial width ($\Delta r$) with distance for the 2007~May~19 event. \emph{Bottom}: Variation in pulse temporal width ($\Delta\tau$) with time. Panels contain data from \emph{STEREO}-A (left) and \emph{STEREO}-B (right). Pulse spatial width here refers to the FWHM of the fitted Gaussian pulse (i.e., $\Delta r = 2\sqrt{2\mathrm{ln}2}\sigma$). The dashed line in all panels indicates the best linear fit to the data. More cases are displayed in the online Figures~\ref{fig:app_broadening_20071207} to \ref{fig:app_broadening_20090213}.}
\label{fig:broadening}
\end{figure*}

\onlfig{13}{
\begin{figure*}
\centerline{
   \includegraphics[width=0.45\textwidth,clip=,trim=0mm 6mm 55mm 93mm]{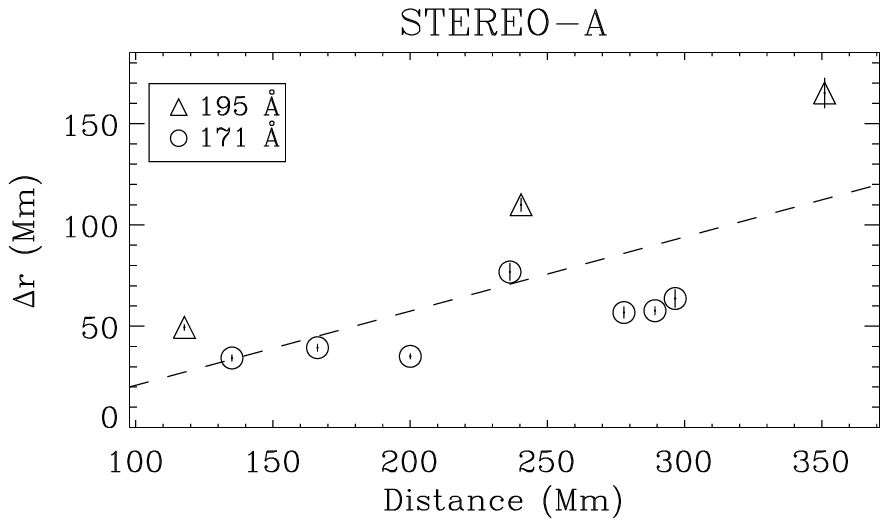}
   \includegraphics[width=0.45\textwidth,clip=,trim=0mm 6mm 55mm 93mm]{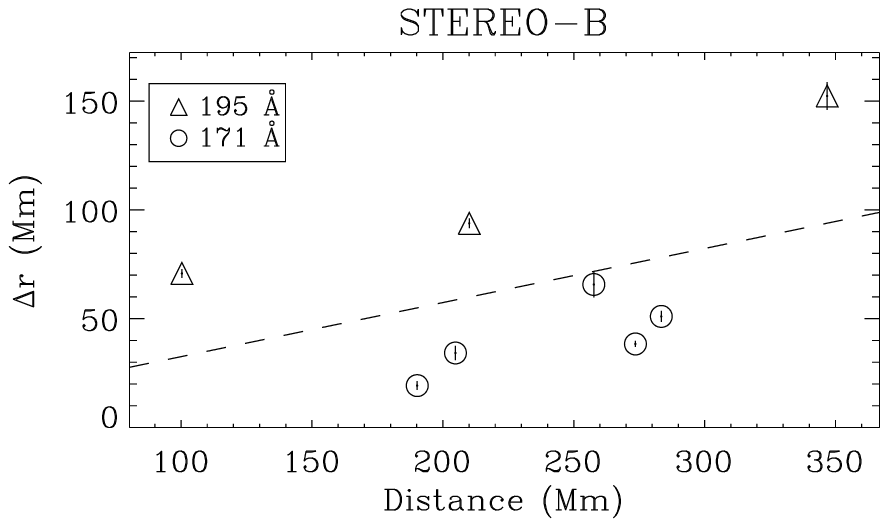}
               }
\centerline{
    \includegraphics[width=0.45\textwidth,clip=,trim=0mm 6mm 55mm 98mm]{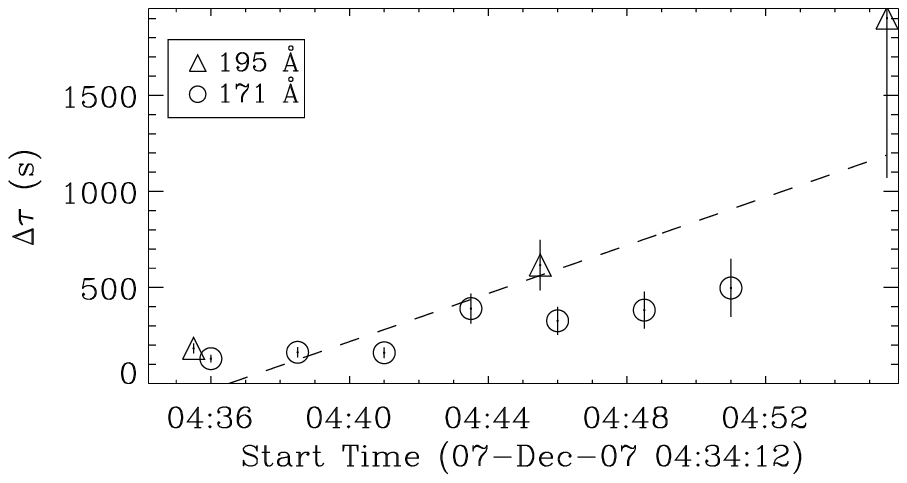}
    \includegraphics[width=0.45\textwidth,clip=,trim=0mm 6mm 55mm 98mm]{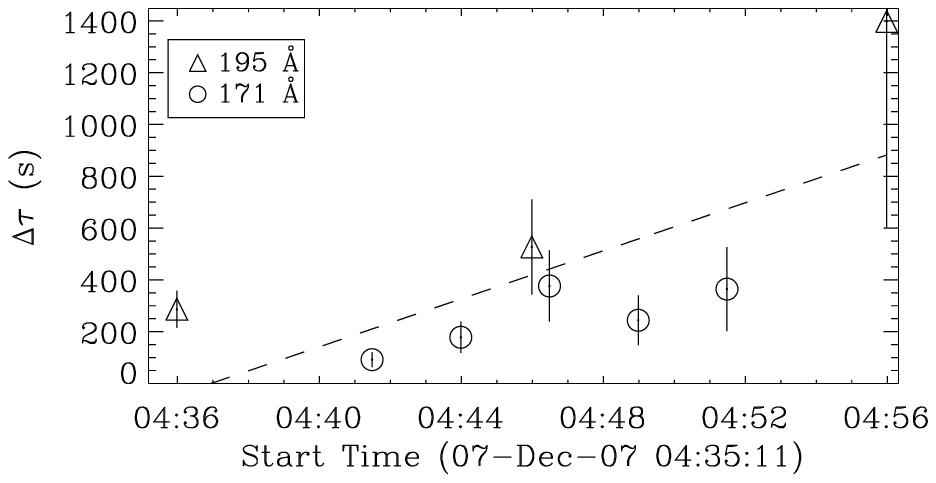}
               }
\caption{Same as for Figure~\ref{fig:broadening} but showing the event on 2007~Dec~07.}
\label{fig:app_broadening_20071207}
\end{figure*}
}
\onlfig{14}{
\begin{figure*}
\centerline{
   \includegraphics[width=0.45\textwidth,clip=,trim=0mm 6mm 55mm 93mm]{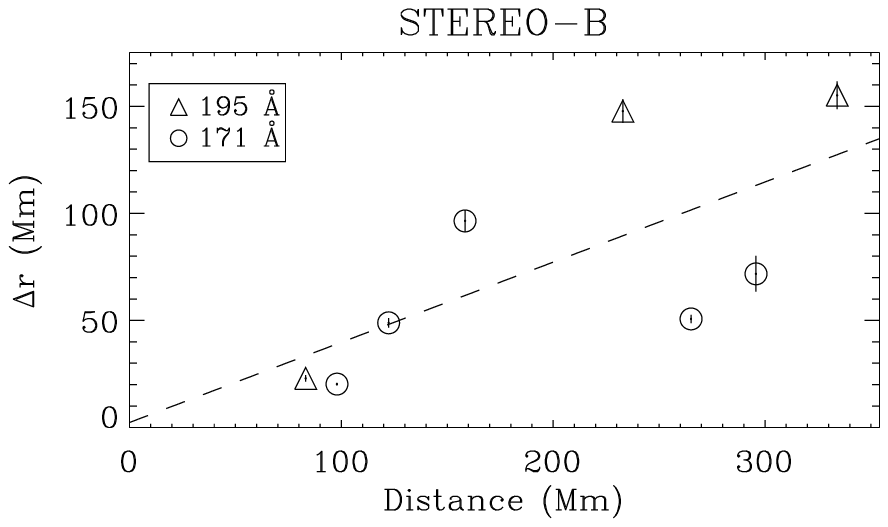}
}
\centerline{
   \includegraphics[width=0.45\textwidth,clip=,trim=0mm 6mm 55mm 98mm]{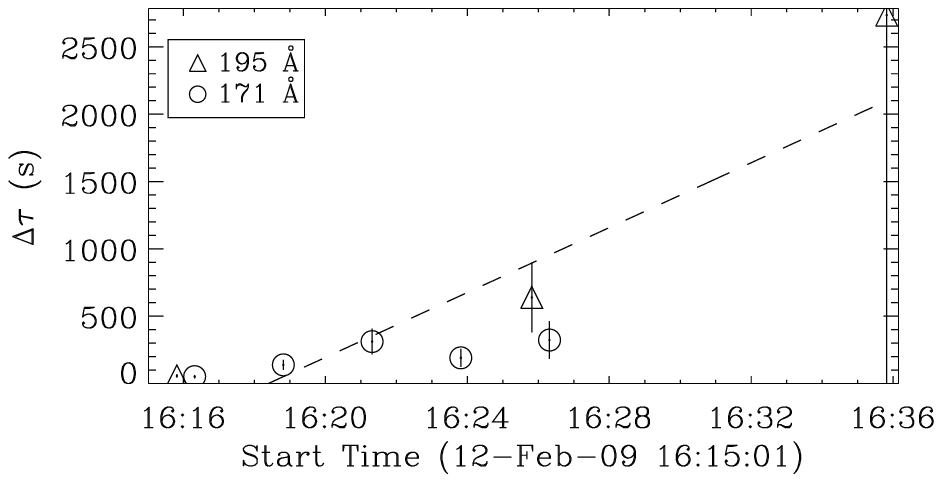}
               }
\caption{Similar to Figure~\ref{fig:broadening}, now only showing \emph{STEREO}-B data for the event on 2009~Feb~12.}
\label{fig:app_broadening_20090212}
\end{figure*}
}
\onlfig{15}{
\begin{figure*}
\centerline{
   \includegraphics[width=0.45\textwidth,clip=,trim=0mm 6mm 55mm 93mm]{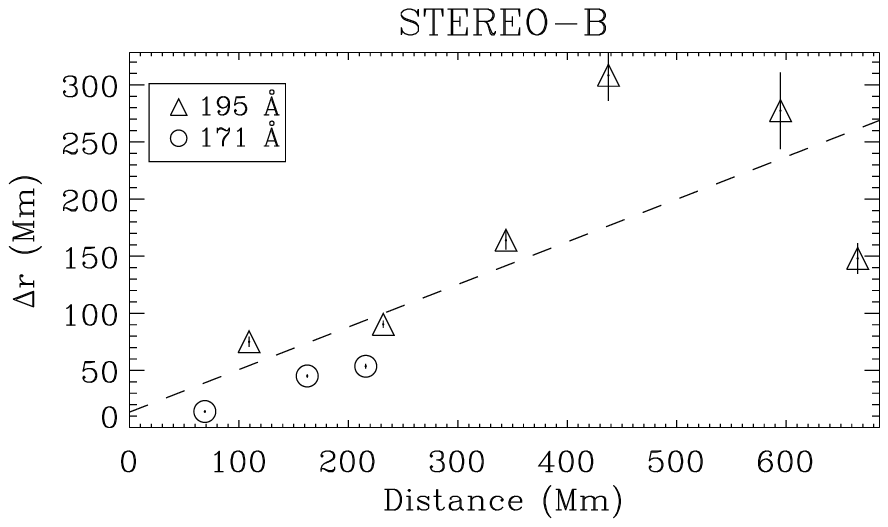}
}
\centerline{
   \includegraphics[width=0.45\textwidth,clip=,trim=0mm 6mm 55mm 98mm]{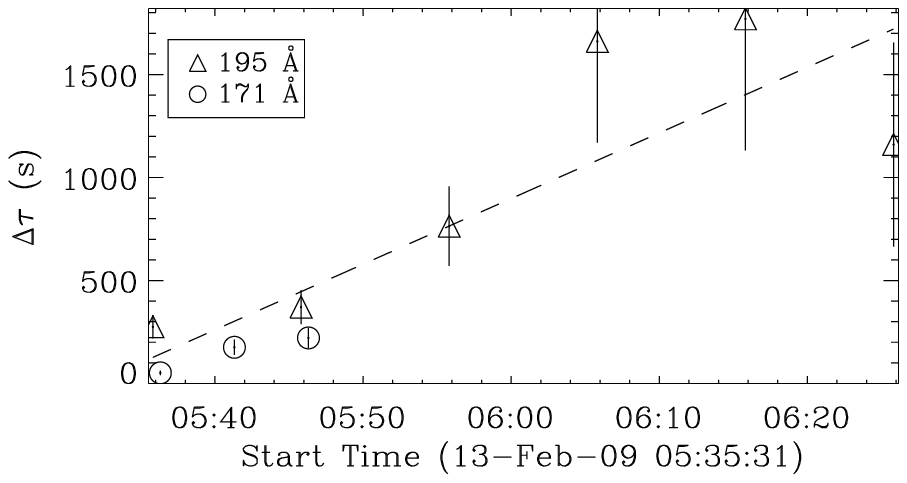}
               }
\caption{Similar to Figure~\ref{fig:broadening}, now only showing \emph{STEREO}-B data for the event on 2009~Feb~13.}
\label{fig:app_broadening_20090213}
\end{figure*}
}

The variation with distance of the full width at half maximum (FWHM; $\Delta r = 2\sqrt{2\mathrm{ln}2}\sigma$) of the Gaussian fit applied to the PBD intensity profile was studied to identify any evidence of pulse broadening. The top two panels in Figure~\ref{fig:broadening} show this variation for \emph{STEREO}-A (left) and \emph{STEREO}-B (right), with measurements from the differing passbands indicated by the different symbols. It is clear from these plots that an increase in pulse width with distance is present for both passbands as observed by both spacecraft. This is indicative of pulse broadening and shows that the CBF spreads out spatially as it propagates.

The variation in the temporal width of the pulse with time was also considered. The temporal width of the pulse, $\Delta\tau$, is defined as,
\begin{equation}\label{eqn:t_width}
\Delta\tau = \frac{\Delta r}{v_{\mathrm{pulse}}},
\end{equation}
where $\Delta r$ is the spatial width of the pulse and $v_{\mathrm{pulse}}$ is the velocity of the pulse. While an increase in the spatial width of the pulse may indicate dispersion, this could be negated by increasing pulse velocity, producing a pulse with constant temporal width. By examining the variation of the temporal width of the pulse with time it is possible to determine if the pulse is indeed dispersive. The resulting plot of the variation in pulse temporal width with time is given in the bottom two panels of Figure~\ref{fig:broadening}. 

The temporal width of the pulse is also observed to increase with time, indicating that the pulse broadens in both space and time with propagation. The increase in spatial and temporal pulse width is apparent in the data from both \emph{STEREO} spacecraft, suggesting that it is a true feature of the disturbance and not an observational artifact. The rate of change of the spatial and temporal width of the pulse with distance and time respectively ($d(\Delta r)/dr$ and $d(\Delta\tau)/dt$) are given in Table~\ref{tbl:events_characteristics}.

The broadening of the CBF pulse in both space and time confirms the previous observations of both \citet{Warmuth:2001ab} and \citet{Veronig:2010ab}. These observations, when taken together with the derived kinematics, are consistent with the interpretation of a CBF as a dispersive, decelerating pulse.

\subsection{Integrated Intensity}
\label{subsect:intensity}

\begin{figure*}
\centerline{
    \includegraphics[width=0.45\textwidth,clip=,trim=0mm 0mm 60mm 0mm]{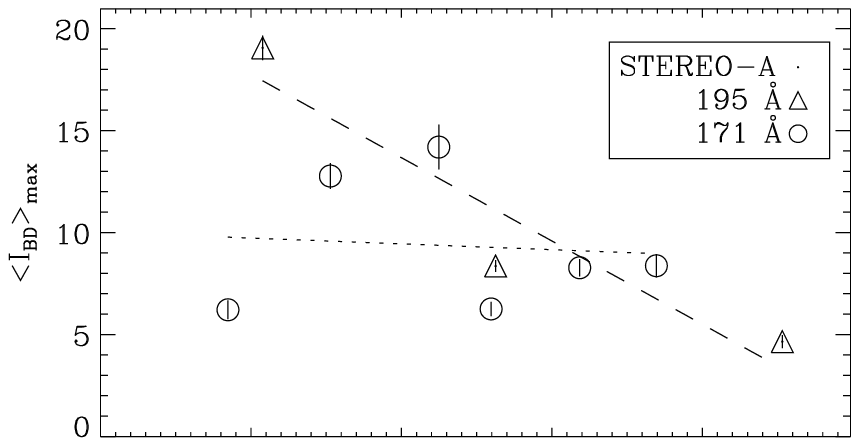}
    \includegraphics[width=0.45\textwidth,clip=,trim=0mm 0mm 60mm 0mm]{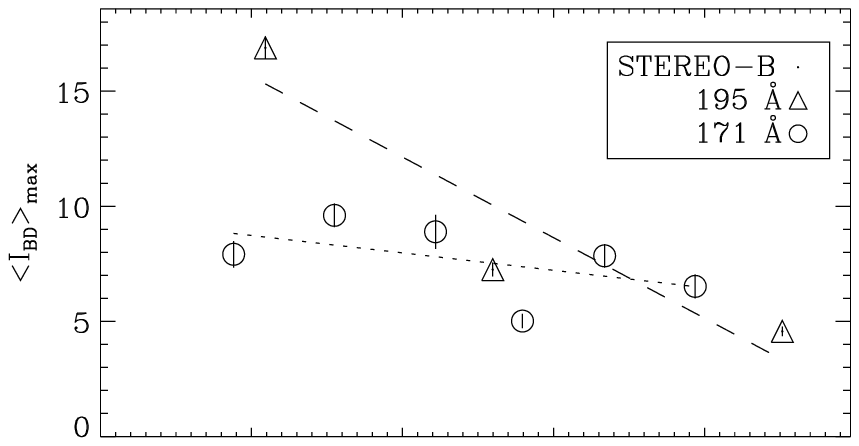}
               }
\caption{\emph{Top}: Variation in peak percentage base difference intensity ($\langle I_{BD}\rangle_{\mathrm{max}}$) with distance. \emph{Middle}: Variation in FWHM of the Gaussian pulse with distance. \emph{Bottom}: Variation in integrated intensity ($I_{tot}$) with distance. Left-hand plots correspond to \emph{STEREO}-A data, with \emph{STEREO}-B data on the right-hand side. The dashed (dotted) line in all panels indicates the best linear fit to the 195~\AA\ (171~\AA) data. More cases are displayed in the online Figures~\ref{fig:app_intensity_20071207} to \ref{fig:app_intensity_200902}.}
\label{fig:intensity}
\end{figure*}

\onlfig{17}{
\begin{figure*}
\centerline{
    \includegraphics[width=0.45\textwidth,clip=,trim=0mm 0mm 60mm 0mm]{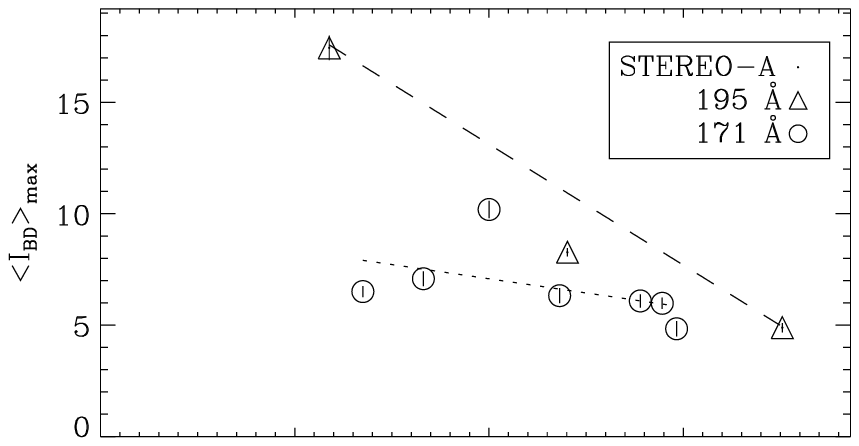}
    \includegraphics[width=0.45\textwidth,clip=,trim=0mm 0mm 60mm 0mm]{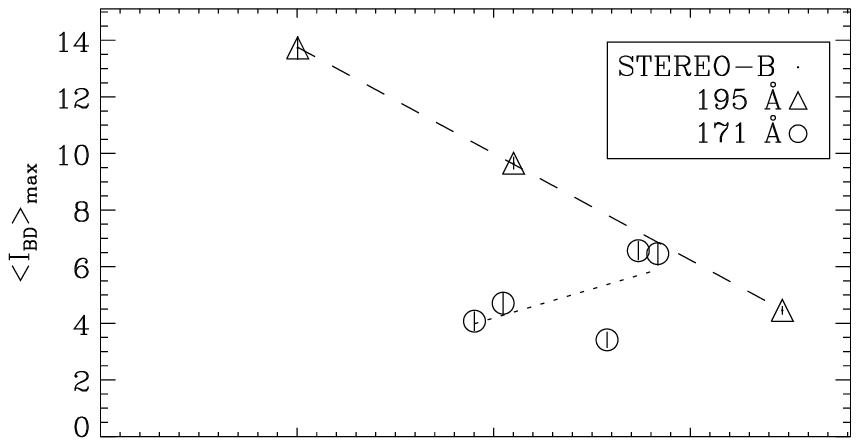}
               }
\caption{Same as for Figure~\ref{fig:intensity} but for the event of 2007~December~07.}
\label{fig:app_intensity_20071207}
\end{figure*}
}
\onlfig{18}{
\begin{figure*}
\centerline{
    \includegraphics[width=0.45\textwidth,clip=,trim=0mm 0mm 60mm 0mm]{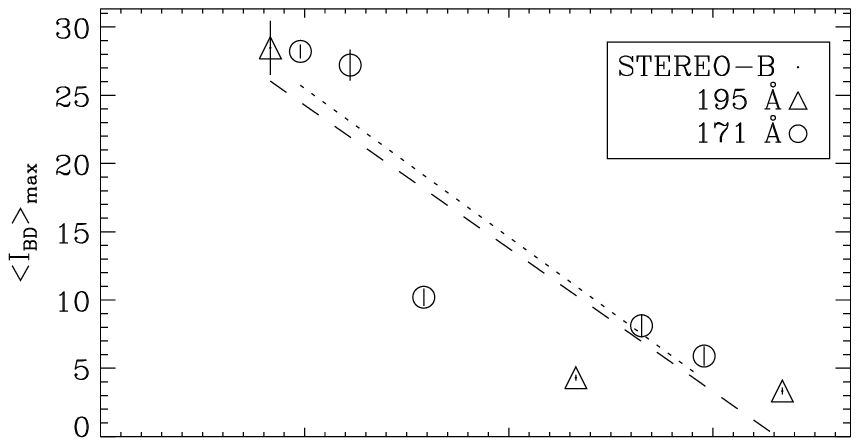}
    \includegraphics[width=0.45\textwidth,clip=,trim=0mm 0mm 60mm 0mm]{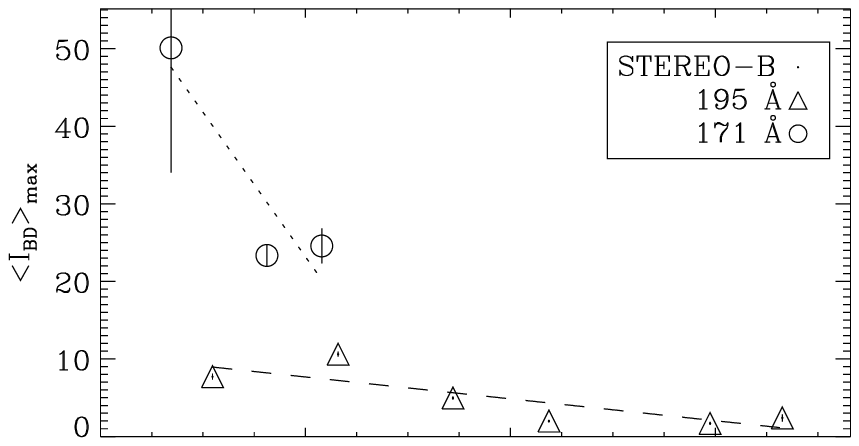}
               }
\caption{Similar to Figure~\ref{fig:intensity} but showing the \emph{STEREO}-B data from the events of 2009~February~12 (left) and 2009~February~13 (right).}
\label{fig:app_intensity_200902}
\end{figure*}
}

The variation in the PA-averaged integrated intensity of the pulse was also examined in order to determine the physical nature of the disturbance. This was found by determining the variation in peak intensity with distance ($\langle I_{BD}\rangle_{\mathrm{max}}$ versus r) and the variation in the FWHM of the pulse with distance ($\Delta r$ versus r). The PA-averaged integrated intensity could then be defined as,
\begin{equation}
I_{\mathrm{tot}} = \Delta r \langle I_{\mathrm{BD}}\rangle_{\mathrm{max}},
\end{equation}
and plotted as a function of distance (see Figure~\ref{fig:intensity} for the resulting plots for the 2007~May~19 event).

This approach was used as the pulse has shown broadening (i.e., dispersion), with the result that the variation in peak intensity with distance may be influenced by this dispersion. The peak intensity of the BD intensity profile was used as this shows the actual emission of the pulse, while the PBD intensity shows the ratio of the pulse emission with respect to the background.

Figure~\ref{fig:intensity} shows the variation in peak pulse intensity, FWHM and PA-averaged BD integrated intensity with distance for each spacecraft. The top plot shows a large discrepancy between the 171~\AA\ and 195~\AA\ observations, with the 195~\AA\ data showing a large decrease with time while the 171~\AA\ data shows no strong variation. The FWHM does follow a linearly increasing trend with distance for both passbands, but the resulting variation in PA-averaged BD integrated intensity is inconclusive in each case. In both cases the 195~\AA\ data appears flat with distance (albeit with some stronger point-to-point variation), while the 171~\AA\ data shows a generally increasing trend. The rate of change of $I_{tot}$ with distance (i.e.,\ d($I_{tot}$)/dr) is given in Table~\ref{tbl:events_characteristics}.

These measurements show that more analysis is required to understand the morphology of the pulse. The higher cadence observations available from \emph{SDO} should allow the variation in peak intensity and PA-averaged integrated intensity with distance to be determined with a much higher degree of accuracy than possible using \emph{STEREO}.

\section{Results}
\label{sect:results}

The intensity profile technique outlined in Section~\ref{sect:observations} is extremely effective at identifying the CBF pulse in observations, with the associated errors much lower than those achieved with previous techniques. Once the sector into which the pulse propagates has been identified, the technique automatically returns the centroid and FWHM of the fitted Gaussian as well as the integrated pulse intensity averaged over position angle at any given image time. The technique also produces accurate and reproducible estimates of the pulse characteristics, making these results more robust. The findings from the event of 2007~May~19 are discussed in detail in Section~\ref{subsect:may19} with several other CBF events summarized in Section~\ref{subsect:other_events}, while all of the results are presented in Table~\ref{tbl:events_characteristics}.

\subsection{2007~May~19 Event}
\label{subsect:may19}

The event of 2007~May~19 displays kinematics consistent with those of a decelerating pulse, with an initial velocity that is towards the upper end of previous estimates. The acceleration term was observed to be negative within the 1-sigma error range, with the similar results from both spacecraft suggesting that these are the true kinematics of the pulse. 

The pulse was observed to display clear spatial and temporal broadening during propagation, indicating that the CBF is dispersive. As a result, the variation with distance of the integrated intensity of the pulse averaged across position angle rather than the peak intensity of the pulse was examined to try and understand the physical nature of the pulse. A decrease in the peak amplitude of a dispersive pulse does not necessarily imply a dissipative pulse (due to the presence of pulse broadening) and may result in a misinterpretation of the true nature of the disturbance.

This event shows inconclusive variation in the PA-averaged integrated pulse intensity with distance from both observed passbands. The higher cadence 171~\AA\ data exhibits a generally increasing trend with distance, while the lower cadence 195~\AA\ observations show negligible variation on average, but strong point-to-point variation. The multi-passband, high-cadence observations afforded by the \emph{Solar Dynamic Observatory} (\emph{SDO}) will allow the true variation (if any) to be determined to a high degree of accuracy.

\subsection{Further CBF Events}
\label{subsect:other_events}

Three additional CBF events from 2007~December~12, 2009~February~12, and 2009~February~13 were also studied using the intensity profile technique (see Figures~\ref{fig:profile_20071207_A} to \ref{fig:profile_20090213_B}). The relationships discussed in Section~\ref{sect:methods} were plotted for these additional events, with approximately similar results observed for each event. Table~\ref{tbl:events_characteristics} shows the kinematics of all the pulses studied and indicates that they displayed similar initial propagation velocities ($\sim$240--450~km~s$^{-1}$). The 2007~May~19 event appears to have been a relatively fast event, with an initial velocity of $\sim$450~km~s$^{-1}$ and a statistically significant non-zero acceleration. The event of 2007~December~07 was much slower, with an initial velocity of $\sim$260~km~s$^{-1}$ and a statistically significant negative acceleration as observed by both \emph{STEREO} spacecraft. The quadrature events of 2009~February~12 and 13 were different from each other despite originating from the same region. The 2009~February~12 event showed a faster initial velocity ($\sim$405~km~s$^{-1}$) and stronger deceleration ($\sim$ $-$291~m~s$^{-2}$), while the 2009~February~13 event had a slower initial velocity ($\sim$274~km~s$^{-1}$) and a much weaker negative acceleration ($\sim$ $-$49~m~s$^{-2}$). The large errors associated with the acceleration terms given here indicate the difficulties associated with accurately determining the kinematics of CBFs from low cadence observations, despite the minimization of errors through the use of both the intensity profile technique and bootstrapping analysis. The distance-time plots for each event are given in the online Figures~\ref{fig:app_kinematics_20071207} and \ref{fig:app_kinematics_200902} for comparison.

\begin{table*}[!t]
\caption{Wave properties of studied CBFs.}
\label{tbl:events_characteristics}
\centering
\begin{tabular}{cccccccc}
\hline \hline
 		&	& \multicolumn{2}{c}{Kinematics} & \multicolumn{2}{c}{Dispersion}  & \multicolumn{2}{c}{Integrated Intensity} \\
Event & Spacecraft & $v_{0}$ & $a$ & $d(\Delta r)/dr$ & $d(\Delta \tau)/dt$ & \multicolumn{2}{c}{$d(I_{tot})/dr$} \\
 		&	& km s$^{-1}$ & m s$^{-2}$ &  &  & 171~\AA\ & 195~\AA\ \\
\hline
2007~May~19	& Ahead 	& $444 \pm 75$	& $-256 \pm 134$ 	& $0.3 \pm 0.1$ 	& $1.0 \pm 0.2$ 	& $1.08 \pm 0.53$ 	& $0.13 \pm 1.07$ \\
 				& Behind 	& $447 \pm 87$	& $-260 \pm 149$ 	& $0.5 \pm 0.1$ 	& $1.1 \pm 0.2$ 	& $1.83 \pm 0.41$ 	& $0.24 \pm 0.61$ \\
2007~Dec~07 	& Ahead 	& $270 \pm 37$ 	& $-152 \pm 73$ 		& $0.4 \pm 0.1$ 	& $1.0 \pm 0.3$ 	& $0.71 \pm 0.37$ 	& $-0.049 \pm 0.002$ \\
				& Behind	& $247 \pm 62$ 	& $-117 \pm 110$ 	& $0.2 \pm 0.2$ 	& $0.8 \pm 0.3$ 	& $2.29 \pm 0.41$ 	& $-1.07 \pm 0.12$ \\
2009~Feb~12 	& Behind 	& $405 \pm 93$   & $-291 \pm 166$ 	& $0.4 \pm 0.2$ 	& $2.0 \pm 0.5$ 	& $-3.27 \pm 2.39$ 	& $-0.59 \pm 0.11$ \\
2009~Feb~13 	& Behind 	& $274 \pm 53$ 	& $-49 \pm 34$ 		& $0.4 \pm 0.1$ 	& $0.5 \pm 0.1$ 	& $4.33 \pm 0.66$ 	& $-0.74 \pm 0.36$ \\
\hline
\end{tabular}
\tablefoot{Kinematic values refer to the mean and standard deviation error of the bootstrapping parameter distributions. Dispersion and integrated intensity values refer to the rate of change of the relevant parameters, resulting from the linear fits shown in Figures~\ref{fig:broadening} and \ref{fig:intensity}.}
\end{table*}

Table~\ref{tbl:events_characteristics} also shows the rate of change of spatial and temporal pulse width $\Delta r$ and $\Delta \tau$ with distance and time respectively, as well as the rate of change of the PA-averaged integrated pulse intensity, $\Delta I_{tot}$, associated with each event (i.e.,\ the slope of the lines shown in Figures~\ref{fig:broadening} and \ref{fig:intensity}). The results indicate that all of the observed events exhibited clear pulse broadening in both the spatial and temporal domains, with the $d(\Delta r)/dr$ and $d(\Delta \tau)/dt$ parameters positive for both the 171~\AA\ and 195~\AA\ passbands. This suggests that pulse broadening is a general characteristic of CBFs and must be accounted for by all theories that seek to explain this phenomenon. The plots showing the variation in both spatial and temporal pulse width with distance and time respectively for each event are given in the online Figures~\ref{fig:app_broadening_20071207} to \ref{fig:app_broadening_20090213}.

The variation in PA-averaged integrated pulse intensity with distance is more difficult to interpret. Both of the online Figures~\ref{fig:app_intensity_20071207} and \ref{fig:app_intensity_200902} show the variation in peak intensity (top), FWHM (middle) and PA-averaged integrated pulse intensity (bottom) with distance. In each case, the 195~\AA\ peak intensity drops with distance while the 171~\AA\ peak intensity variation is inconclusive, although the FWHM of each passband tends to increase. For the 2007~December~07 event, the resulting PA-averaged integrated pulse intensity shows a slight decrease with distance for the 195~\AA\ data, while showing an apparent increase with distance for the 171~\AA\ data, with a large separation between the 171~\AA\ and 195~\AA\ passbands. In contrast, both the 171~\AA\ and 195~\AA\ observations tend to drop with distance for the 2009~February~12 event, while the 2009~February~13 event shows an increase and decrease with distance for the 171~\AA\ and 195~\AA\ data respectively. These observations suggest that the variation in PA-averaged integrated pulse intensity with distance is not well-defined and requires further investigation.

\section{Conclusions}
\label{sect:conclusion}

Several CBFs were studied by applying a semi-automated technique to data from the EUVI telescopes onboard both \emph{STEREO} spacecraft. This technique applies Gaussian fits to intensity profiles across propagating CBFs, returning the position, width, and PA-averaged integrated intensity of the pulse for each observation. This intensity profile approach is similar to CBF identification techniques previously proposed by \citet{Warmuth:2004ab,podladchikova2005,Wills-Davey:2006ab,Veronig:2010ab}. However, by combining it with a statistically rigorous bootstrapping method and the high cadence observations afforded by \emph{STEREO}/EUVI we can minimise the errors typically encountered with the analysis of CBFs.

The similarity in derived velocity between this work and previous investigations is interesting given that most previous works have used point-and-click techniques applied to running-difference images. These studies identify the forward edge of the CBF at a given time, which is then used to determine the kinematics of the disturbance as a whole. The analyses performed using such techniques have mainly returned kinematics that suggest a zero acceleration (i.e., constant velocity) interpretation of the CBF phenomenon. In contrast, our semi-automated technique uses the pulse centroid derived from percentage base difference intensity profiles to return a constant, non-zero acceleration (i.e., variable velocity) interpretation. In the presence of pulse broadening, as found here, the forward edge of a decelerating pulse will appear to move faster than the pulse centroid. This suggests that the true kinematics of CBFs may have previously been disguised through use of the wrong position within the pulse profile to characterise its location. We also note that a variable acceleration may be present, but this can only be studied with an increased number of data points.

A positive increase in the variation of the width of the Gaussian fit to the CBF pulse in both the spatial and temporal domains was observed for all events studied. While the increase in spatial pulse width is suggestive of a dispersive pulse, this is confirmed by the observed increase in the temporal width of the pulse. From Equation~\ref{eqn:t_width}, an increase in the spatial pulse width may be negated by an increase in pulse velocity, producing a pulse with a constant temporal width. However, Figures~\ref{fig:broadening} to \ref{fig:app_broadening_20090213} show that this is not the case. This indicates that all of the CBFs studied showed significant pulse broadening and may be interpreted as being dispersive. Although some previous studies have suggested a dispersive nature for ``EIT waves'', the true extent of this dispersion may have been disguised by the nature of point-and-click analyses of running-difference images. 

The variation in PA-averaged integrated pulse intensity with distance was studied rather than the variation in peak pulse intensity due to the presence of pulse broadening. This was found by multiplying the FWHM by the peak pulse intensity at a given time, and produced inconclusive results. While the peak intensity of the pulse was generally observed to decrease with distance and the FWHM was observed to increase with distance in both passbands, the PA-averaged integrated intensity typically showed no strong variation with distance in either passband, although a strong offset between the 171~\AA\ and 195~\AA\ data was typically observed. Contrasting results were found for the 2009~February~12 and 2009~February~13 events respectively, suggesting that further investigation using simultaneous analysis of multiple passbands at very high cadence is required, something that will be routinely available from the \emph{SDO} spacecraft.

The results of our analysis suggest that the studied CBFs may be interpreted as dispersive pulses exhibiting negative acceleration. This is consistent with the fast-mode magnetoacoustic wave interpretation for a freely-propagating ``EIT wave''. The variation in the integrated intensity of the pulse was inconclusive, implying that more analysis is required to definitively determine the physical nature of the disturbance.

The consistency of the results between the events studied suggests that the conclusions drawn here may be applicable to a larger sample of CBFs. The initial velocities of the CBFs are comparable to the lower range of estimated Alfv\'{e}n speeds proposed by \citet{Wills-Davey:2007oa}, suggesting that the randomly structured nature of the coronal magnetic field may have an important effect on the propagation of CBFs; an effect shown in simulations performed by \citet{Murawski:2001ab}. Although the kinematics and dispersive nature of the pulses suggest a magnetoacoustic wave interpretation may be correct, further study is required. The $\sim$10~s temporal cadence across multiple EUV passbands available with the launch of the \emph{SDO} spacecraft should allow these results to be studied in much greater detail.

\begin{acknowledgements}
D.M.L. is a Government of Ireland Scholar supported by the Irish Research Council for Science, Engineering and Technology (IRCSET), R.T.J.McA. carried out this work as a Marie Curie Fellow under FP6, and D.S.B. is a Marie Curie Fellow funded under FP7. We would like to thank the \emph{STEREO}/SECCHI consortium for providing open access to their data and technical support as well as A.~M.~Veronig and I.~W.~Kienreich for useful discussions regarding this subject. We would also like to thank the anonymous referee for many useful comments which strengthened this paper.
\end{acknowledgements}

\end{document}